\documentclass[]{ws-p9-75x6-50} 

\begin{document}   

\title{Covariant Geometric Prequantization of Fields } 

\author{Igor V. Kanatchikov}

\address{Institute of Theoretical Physics \\
Friedrich Schiller University,   
07743 Jena, Germany \\
 Tallinn Technical University, 19086 Tallinn, Estonia  
}

\maketitle

\abstracts{
{A~geometric~prequantization~formula~for~the 
Poisson-Gerstenhaber~bracket~of~forms}  
{found~within~the 
De~Donder-Weyl~Hamiltonian~formalism~earlier~is~presented.~The} 
related aspects of covariant geometric quantization  
of field theories are sketched. }  

\renewcommand{\rightnote}{\small submitted to: 
{\em Proc. Ninth Marcel Grossmann Meeting, 
Roma (Italy), July 2-8, 2000} }


\newcommand{\beq}{\begin{equation}}
\newcommand{\eeq}{\end{equation}}
\newcommand{\beqa}{\begin{eqnarray}}
\newcommand{\eeqa}{\end{eqnarray}}
\newcommand{\nn}{\nonumber}

\newcommand{\half}{\frac{1}{2}}

\newcommand{\xt}{\tilde{X}}

\newcommand{\uind}[2]{^{#1_1 \, ... \, #1_{#2}} }
\newcommand{\lind}[2]{_{#1_1 \, ... \, #1_{#2}} }
\newcommand{\com}[2]{[#1,#2]_{-}} 
\newcommand{\acom}[2]{[#1,#2]_{+}} 
\newcommand{\compm}[2]{[#1,#2]_{\pm}}

\newcommand{\lie}[1]{\pounds_{{\mbox{\footnotesize${#1}$}}}}
\newcommand{\co}{\circ}
\newcommand{\sgn}[1]{(-1)^{#1}}
\newcommand{\lbr}[2]{ [ \hspace*{-1.5pt} [ #1 , #2 ] \hspace*{-1.5pt} ] }
\newcommand{\lbrpm}[2]{ [ \hspace*{-1.5pt} [ #1 , #2 ] \hspace*{-1.5pt}
 ]_{\pm} }
\newcommand{\lbrp}[2]{ [ \hspace*{-1.5pt} [ #1 , #2 ] \hspace*{-1.5pt} ]_+ }
\newcommand{\lbrm}[2]{ [ \hspace*{-1.5pt} [ #1 , #2 ] \hspace*{-1.5pt} ]_- }
\newcommand{\pbr}[2]{ \{ \hspace*{-2.2pt} [ #1 , #2\hspace*{1.5 pt} ] 
\hspace*{-2.7pt} \} }
\newcommand{\we}{\wedge}
\newcommand{\dv}{d^V}
\newcommand{\nbrpq}[2]{\nbr{\xxi{#1}{1}}{\xxi{#2}{2}}}
\newcommand{\lieni}[2]{$\pounds$${}_{\stackrel{#1}{X}_{#2}}$  }

\newcommand{\rbox}[2]{\raisebox{#1}{#2}}
\newcommand{\xx}[1]{\raisebox{1pt}{$\stackrel{#1}{X}$}}
\newcommand{\xxi}[2]{\raisebox{1pt}{$\stackrel{#1}{X}$$_{#2}$}}
\newcommand{\ff}[1]{\raisebox{1pt}{$\stackrel{#1}{F}$}}
\newcommand{\dd}[1]{\raisebox{1pt}{$\stackrel{#1}{D}$}}
\newcommand{\nbr}[2]{{\bf[}#1 , #2{\bf ]}}
\newcommand{\der}{\partial}
\newcommand{\oo}{$\Omega$}
\newcommand{\Om}{\Omega}
\newcommand{\om}{\omega}
\newcommand{\eps}{\epsilon}
\newcommand{\si}{\sigma}
\newcommand{\Lm}{\bigwedge^*}

\newcommand{\inn}{\hspace*{2pt}\raisebox{-1pt}{\rule{6pt}{.3pt}\hspace*
{0pt}\rule{.3pt}{8pt}\hspace*{3pt}}}
\newcommand{\sro}{Schr\"{o}dinger\ }
\newcommand{\bm}{\boldmath}
\newcommand{\vol}{\omega}
               \newcommand{\dvol}[1]{\der_{#1}\inn \vol}

\newcommand{\bd}{\mbox{\bf d}}
\newcommand{\bder}{\mbox{\bm $\der$}}
\newcommand{\bI}{\mbox{\bm $I$}}

\newcommand{\ga}{\gamma} 
\newcommand{\de}{\delta} 
\newcommand{\Ga}{\Gamma} 
\newcommand{\gmu}{\gamma^\mu}
\newcommand{\gnu}{\gamma^\nu}
\newcommand{\ka}{\kappa}
\newcommand{\hka}{\hbar \kappa}
\newcommand{\lapl}{\bigtriangleup}
\newcommand{\psib}{\overline{\psi}}
\newcommand{\Psib}{\overline{\Psi}}
\newcommand{\derts}{\stackrel{\leftrightarrow}{\der}}
\newcommand{\what}[1]{\widehat{#1}}

\newcommand{\bx}{{\bf x}}
\newcommand{\bk}{{\bf k}}
\newcommand{\bq}{{\bf q}}

\newcommand{\omk}{\omega_{\bf k}} 
\newcommand{\lpl}{\ell}
\newcommand{\zb}{\overline{z}} 

\newcommand{\BPsi}{\Bbb \Psi} 
\newcommand{\BH}{\Bbb H} 
\newcommand{\BS}{\Bbb S} 
\newcommand{\BN}{\Bbb N} 


\newcommand{\introA}{ 
The goal of geometric quantization \cite{woodh,simms,roman-roy} 
is to construct quantum objects 
starting from the geometry of corresponding classical objects. 
It clarifies the origin and the essence 
of the quantization procedure. 
 
The use of geometric quantization in field theory 
\cite{} is 
hampered by the infinite dimensional character of the 
structures of the commonly used canonical 
Hamiltonian formalism in field  theory. It is often difficult 
to give to the formulas formally generalized to infinite 
dimensions a proper mathematical foundation. 
Moreover, the framework implies the procedure of the space+time 
splitting which restricts it to globally hyperbolic space-times. 
It seems unlikely 
that this restriction is proper to 
the problem of quantization of general relativity. 
Some conceptual and technical problems in quantum gravity, 
such as ``the problem of time''  and the meaning, both 
mathematical and physical, of the Wheeler-DeWitt equation, 
can be traced back up to the features of the canonical formalism 
in question.  

We put forward another approach to field quantization 
which emanates from an alternative generalization 
of the Hamiltonian formalism to field theory which 
neither needs the space+time splitting,  nor an infinite dimensional 
geometric framework: 
given a Lagrangian 
...
} 


\noindent

The use of the methods of geometric quantization\cite{gq} 
in field theory 
is  
severely limited by the infinite dimensionality 
of the standard field theoretic canonical Hamiltonian formalism  
because of the difficulties of substantiation of 
formal generalizations  
of geometric constructions to infinite dimensions. 

We suggest an approach based 
on the {\em De Donder-Weyl} (DW) Hamiltonian form of field equations 
\vspace*{-7pt} 
\beq 
\der_\mu \phi^a (x) = {\der H}/{\der p^\mu_a} , 
\quad 
\der_\mu  p^\mu_a (x) = {\der H}/{\der \phi^a }  , 
\eeq 
where 
 $p_a^\mu:=\der L / \der \phi^a_{,\mu}$,  
$H(\phi^a,p^\mu_a,x^\nu):= \phi^a_{,\mu} p^\mu_a -L$, 
$L(\phi^a, \phi^a_{,\mu}, x^\nu)$ is   a Lagrangian density,    
which incorporates the field dynamics into a finite dimensional 
{\em polymomentum phase space}: $(p^\mu_a, \phi^a, x^\mu)$ 
and requires no space+time splitting. 
The Poisson bracket for the above 
DW formulation is defined on differential forms 
which represent dynamical variables\cite{ikanat,ik2}. The basic structure  
is the {\em polysymplectic form }    
$\Omega := d\phi^a\we dp^\mu_a \we \omega_\mu$  
($\omega_\mu:= \der_\mu\inn (dx^1\we ...\we dx^n)$)  
which maps (``horizontal'') 
 $f$-forms  $F$
to (``vertical'') multivectors $X_F$ of degree $(n-f)$:
$
X_F\inn\Omega = dF. 
$
The map exists for the so-called {\em Hamiltonian forms} 
the space of which is closed 
with respect to the {\em co-exterior product}\cite{ik2}: 
$F\!\bullet G :=*^{-1}(*F\!\we\! *G)$. 
The bracket 
\beq
\pbr{F}{G} := (-)^{n-f} \lie{X_F}(G),  
\eeq 
where 
$\lie{X_F} := [X_F,d]= X_F\co d - (-)^{n-f}d\co X_F$, 
equips 
the space of Hamiltonian forms 
with a {\em Gerstenhaber algebra}
structure\cite{ikanat,ik2,paufler}:  
\beqa
&&\pbr{F}{G} = -(-1)^{g_1 g_2}
\pbr{G}{F}, \nn \\ 
\mbox{$(-1)^{g_1 g_3} \pbr{F}{\pbr{G}{K}}$} 
&\!+\!& 
\mbox{$(-1)^{g_1 g_2} \pbr{G}{\pbr{K}{F}}$} 
+ 
\mbox{$(-1)^{g_2 g_3} \pbr{K}{\pbr{F}{G}}\!=\! 0,$}  \\
\pbr{F}{G\bullet K} 
&=&
\pbr{F}{G}\bullet K 
+ (-1)^{g_1(g_2+1)}G\bullet\pbr{F}{K}, \nn 
\eeqa 
where $g_1 = n-f-1$,  $g_2 = n-g-1$,  $g_3 = n-k-1$. 

How to quantize fields using the above generalization of 
Poisson brackets? 
Our recent work 
on ``precanonical 
quantization''\cite{ikanat-quant}${}^{-}$\cite{ig5} 
 based on  heuristic quantization of a small subalgebra 
of (3) 
leads to a  
generalization of quantum 
theoretic formalism in which the polymomenta operators 
are $\hat{p}_a^\mu=-i\hbar\kappa\gamma^\mu \frac{\der}{\der \phi^a}$ 
and the covariant 
``multi-temporal'' analogue of the Schr\"odinger equation 
for 
$\Psi=\Psi(\phi^a,x^\mu)$ reads  
\beq
i\hbar\kappa \gamma^\mu\der_\mu \Psi = \what{H}\Psi, 
\eeq
where the constant 
$\kappa\! \sim\! \frac{1}{[{\tt length}]^{(n-1)}}$ 
is of the ultra-violet cutoff scale and,  
for the scalar field, 
$\what{H}=-\half \hbar^2\kappa^2 \der_{\phi\phi}^2 
+ V(\phi)$. 
 It is notable that 
Eq.~(4) enables us to derive the standard functional differential 
Schr\"odinger equation in quantum field theory\cite{ik-schr}.

\newcommand{\oldtexti}{
where for the scalar field 
$\what{H}=-\half \hbar^2\kappa^2 \der_{\phi\phi}^2 
+ V(\phi)$; 
the constant 
$\kappa\! \sim\! \frac{1}{[{\tt length}]^{(n-1)}}$ 
is of the ultra-violet cutoff scale.  
 It is notable that 
the standard functional differential 
Schr\"odinger equation 
can be derived from Eq.~(4)\cite{ik-schr}.  
} 


The main result of the present contribution is 
a generalization of a 
cornerstone  of geometric quantization, 
the prequantization formula,  to 
the present framework\cite{ik-gq}: the operator 
\vspace*{-10pt} 
\beq
 \hat{F}= i\hbar \pounds_{X_F} 
+ (X_F{} \inn \Theta) \bullet + F\bullet , 
\eeq 
with $d\Theta:=\Omega$,  
is a prequantization of a Hamiltonian form $F$ 
in the sense that  
\beq
[ \hat{F_1}, \hat{F_2}] = - i\hbar \what{\pbr{F_1}{F_2}} ,
\eeq 
where $[A,B] = A\co B - (-1)^{\deg A \deg B} B\co A $ 
is a graded commutator. The operator (5) is nonhomogeneous: 
the degree of the first term is $(n-f-1)$ 
while the degree of the other two is $(n-f)$.
Therefore, the first two terms can be viewed as a covariant 
derivative $\nabla_{X_F}$  corresponding to a 
superconnection\cite{quillen}; then the polysymplectic form 
can be viewed as the curvature of the superconnection. 
Prequantum operators (5) can act on prequantum Hilbert space of 
nonhomogeneous differential forms 
$\Psi(\phi^a, p_a^\nu, x^\nu) = 
\sum_{p=0}^n \psi_{\mu_1...\mu_p} dx^{\mu_1}\we ...\we dx^{\mu_p}$  
 with the scalar product 
$\left < \Psi,\Psi  \right > (x) 
= 
\sum_{p=0}^n \int (\psib_{\mu_1...\mu_p}\psi^{\mu_1...\mu_p}){\rm Vol}$,
where 
Vol$:=\prod_{a=1}^m d\phi^a \we dp_a^1 \we dp_a^2 \we ... \we dp_a^n$ 
(no summation over index $a$ here!).  
However, it is more suitable to 
 Cliffordize the above expressions according to the rule 
$\omega_\mu\bullet =  \kappa^{-1}\gamma_\mu$. Then the Cliffordized 
prequantum operators (5) can act on spinor wave functions 
$\Psi(\phi^a, p_a^\nu, x^\nu)$ with the positive definite scalar 
product $\left < \Psi,\Psi  \right > = \int\int_\sigma \Psib\gamma^\mu\Psi$ 
Vol$\we \omega_\mu$, where $\sigma$ is 
a space-like hypersurface. Using the ``vertical 
polarization'' 
with $\Psi=\Psi(\phi^a, x^\nu)$ 
one can derive from (5) the operators 
already known in precanonical 
quantization\cite{ikanat-quant}${}^{-}$\cite{ig5}.  
Thus, the structures of DW theory\cite{ikanat,ik2} 
naturally lead to the 
notions of 
Clifford/spinor 
bundles and superconnections\cite{quillen} 
as a framework of generalizing the techniques of geometric 
quantization\cite{gq} 
to field theory, 
 requiring neither a space+time splitting nor 
an infinite dimensional geometry.


\end{document}